\documentclass[twocolumn,english,aps,prb,superscriptaddress]{revtex4}
\usepackage[T1]{fontenc}
\usepackage[latin1]{inputenc}
\usepackage{graphicx}
\usepackage{amssymb}
\usepackage{babel}


\begin{document}

\title{Realistic quantum manipulation of two-level system fluctuators}

\author{L. Tian}

\affiliation{School of Natural Sciences, University of California, Merced, CA
95344 USA}

\email{ltian@ucmerced.edu}

\author{K. Jacobs}

\affiliation{Department of Physics, University of Massachusetts at Boston, 100
Morrissey Blvd, Boston, MA 02125 USA}

\email{kjacobs@cs.umb.edu}

\date{\today}

\begin{abstract}
Two-level system fluctuators in superconducting devices have demonstrated coherent coupling with superconducting qubits. Here, we show that universal quantum logic gates can be realized in these two-level systems solely by tuning a superconducting resonator in which they are imbedded. Because of the large energy separation between the fluctuators, conventional gate schemes in the cavity QED approach that are widely used for solid-state qubits cannot be directly applied to the fluctuators. We study a scheme to perform the gate operations by exploiting the controllability of the superconducting resonator with realistic parameters. Numerical simulation that takes into account the decay of the resonator mode shows that the quantum logic gates can be realized with high fidelity at moderate resonator decay rate.  The quantum logic gates can also be realized between fluctuators inside different Josephson junctions that are connected by a superconducting loop. Our scheme can be applied to explore the coupling between two-level system fluctuators and superconducting resonators as well as the coherent properties of the fluctuators.

\end{abstract}


\maketitle

\section{introduction}
Spurious two-level system (TLS) fluctuators are considered a serious source of low-frequency noise in superconducting qubits~\cite{MakhlinRMP}, and the characterization of these fluctuators in solid-state devices has a long history~\cite{DuttaRMP}.
Most recently, coherent coupling between TLS fluctuators and a superconducting phase qubit was observed via the novel energy splittings in spectroscopic measurements~\cite{SimmondsPRL,KimRecentExp,MartinPRL}. It was shown that the TLS fluctuators have much longer decoherence times than the superconducting qubits, raising the possibility of realizing quantum manipulation on these fluctuators~\cite{ZagoskinPRL}. 

The key question in manipulating the TLS fluctuators is how to implement the required coherent manipulation and readout. Located sparsely inside solid-state devices,  the  fluctuators usually do not interact with each other, and their states are hard to control. The coupling between the fluctuators and solid-state devices provides us with a tool to achieve the quantum manipulation~\cite{SimmondsPRL,MartinPRL}. However, conventional gate schems using cavity QED approach that are usually exploited for solid-state qubits cannot be applied to this system because of the large energy separation between the fluctuators. In this work, we will present a gate scheme that exploits the controllability of the superconducting resonator to implement high fidelity gates on the TLS fluctuators~\cite{BlaisPRA2004, MigliorePRB, ValenzuelaScience, ChiorescuNature, KochPRL2006}, even when the decay of the resonator is a few megahertz.  The superconducting resonator acts as a knob that controls the dynamics of individual  fluctuator, as well as coupling them together. Working with practical parameters from the superconducting Josephson junction resonator, we will design single-qubit and two-qubit quantum logic gates in the presence of resonator decay. Our scheme takes into account the full coupling Hamiltonian between the  TLS fluctuators and the resonator. 
Readout of the  fluctuators can also be performed by measuring the transmission through the resonator. This scheme can be extended to  fluctuators in different Josephson junctions by connecting  the junctions into the same superconducting loop due to the nonlocal nature of the microwave mode of the resonator.  This work hence provides a realizable design for coherent manipulation of multiple TLS  fluctuators, which is closely related to current experimental efforts  in studying the fluctuators and their coupling with superconducting resonator modes. 

Various superconducting resonators in the microwave regime, including superconducting transmission lines, Josephson junctions, SQUID's, and superconducting lumped element resonators, have recently been demonstrated and have shown quantum behavior
and strong coupling with superconducting qubits~\cite{SillanpaaNature2007, MajerNature2007, Osborn2007, HouckNature2007}. Superconducting resonators are also promising systems for studying quantum effects such as single photon generation and lasing~\cite{photonNakamura}, and one of us has shown recently that a Josephson junction can be used to probe various properties of TLS  fluctuators, e.g. to resolve the mechanism that couples the  fluctuators to the junction~\cite{TianPRL2007}. While we will focus on the Josephson junction resonator, we want to emphasize that our results can be readily generalized to other superconducting resonators~\cite{Osborn2007,Cicak2008}. The paper is organized as the following. In Sec.~\ref{system}, we will study the coupled system of  the fluctuators and a Josephson junction resonator, including the driving on the resonator. In Sec.~\ref{dispersive}, we will derive the effective Hamiltonian for the  TLS fluctuators in the dispersive regime where the quantum operations are implemented. We will also derive the residual coupling between the  fluctuators  and the resonator in this regime. In Sec.~\ref{gates}, we will present detailed scheme for single-qubit and two-qubit quantum logic gates.  Then, we will estimate the decoherence of the  fluctuators during the gate operations in Sec.~\ref{decoherence}.  We will also test the fidelity of the quantum operations with numerical simulation of the full Hamiltonian, taking the resonator decay into account.  In Sec.~\ref{discussions}, we will discuss the readout of the fluctuators and the extension of gate scheme to fluctuators inside different junctions. The conclusions will be given in Sec.~\ref{conclusions}.

\section{The System\label{system}}
\begin{figure}
\includegraphics[width=8cm,clip]{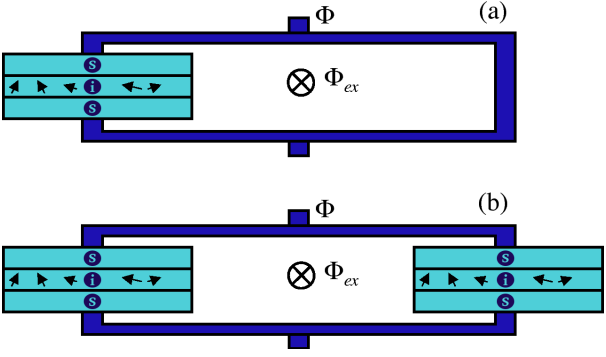}
\caption{\label{fig1} A Josephson junction resonator containing spurious two-level system fluctuators denoted by arrows. (a)  Fluctuators in a single junction and (b)  fluctuators in different junctions. }
\end{figure}

Consider the system in Fig.~\ref{fig1} (a), where
TLS  fluctuators inside the amorphous layer of a Josephson junction couple with
the junction resonator in an RF SQUID loop. With total capacitance
$C_{0}$, Josephson energy $E_{J}$, loop inductance $L$, and magnetic
flux $\Phi_{ex}$ inside the SQUID loop, the Hamiltonian of the
resonator can be written as 
\begin{equation}
H_{c}=\frac{P_{\Phi}^{2}}{2C_0}-E_{J}\cos(2e\Phi/\hbar)+\frac{(\Phi+\Phi_{ex})^{2}}{2L}
\end{equation}
in terms of the phase $\Phi$ and the conjugate momentum
$P_{\Phi}$. This Hamiltonian can be approximated as an oscillator mode
with a phase shift $\Phi_{s}$ from the origin:
\begin{equation}
H_{c} \approx P_{\Phi}^{2}/(2C_{0})+C_{0}\omega_{c}^{2}(\Phi-\Phi_{s})^{2}/2,
\end{equation} 
and  the phase shift satisfies 
\begin{equation}
\hbar \Phi_{s} + 2eL E_{J}\sin(2e\Phi_s/\hbar) = - \hbar \Phi_{ex}. 
\end{equation} 
The frequency of the resonator can be written as~\cite{OrlandoBook1991}
\begin{equation}
    \omega_{c}= \sqrt{ \frac{1}{LC_{0}}+\frac{4e^2E_{J}\cos(2e\Phi_s/\hbar)}{\hbar^2C_{0}}},
\end{equation}
which can be tuned in a large range by the magnetic flux $\Phi_{ex}$.  In addition, driving can be applied to the resonator by e.g. applying an external radio-frequency current $\delta I_{c}$ to the resonator with $\delta I_{c}\Phi$. 

The TLS  fluctuators reside inside the tunneling layer and can couple with the junction resonator by various mechanisms. For example, the coupling to the critical current of the junction takes the form $ -(2e/\hbar)E_{J}\Phi\sum_{n}\vec{j}_{n}\cdot\vec{\sigma}_{n}$, where $\vec{j}_{n}$ is the polarization and magnitude of the coupling. Denoting the resonator annihilation operator by $a$ with
$\Phi-\Phi_{s}=\sqrt{\hbar/(2C_{0}\omega_{c})}(a+a^{\dagger})$.
Let  $\vec{j}_{n}=(j_{xn},0,0)$ for simplicity and $\omega_{d}$ be the driving frequency. The total Hamiltonian of the coupled system in the rotating frame can be written as 
\begin{eqnarray}
H_{t} &=& H_{c}+H_{1}+H_{\kappa} \label{Ht} \\ 
H_{c} &=& \hbar\Delta_{c}a^{\dagger}a+\epsilon(a+a^{\dagger}) \\
H_{1} &=& \sum_{n}\left[(\hbar\Delta_{n}/2)\sigma_{nz}+g_{n}(a\sigma_{n+}+a^{\dagger}\sigma_{n-})\right] \\
H_{\kappa} &=& \sum_{k}\hbar\omega_{k}a_{k}^{\dagger}a_{k}+c_{k}(a_{k}^{\dagger}a+a^{\dagger}a_{k}) 
\end{eqnarray}
where $H_{c}$ is the Hamiltonian of the driven resonator mode with the detuning $\Delta_{c}=\omega_{c}-\omega_{d}$ and the driving amplitude $\epsilon=\delta I_{c}\sqrt{\hbar/(2C_{0}\omega_{c})}$,  $H_{1}$ is the Hamiltonian of the  fluctuators including the coupling between the  fluctuators and the resonator mode, and $H_{\kappa}$ is  the Hamiltonian of the thermal bath connected to the resonator. Here, the index $n$ labels different fluctuators, $\sigma_{n\alpha}$ are the Pauli operators, $\Delta_{n}=\omega_{n}-\omega_{d}$ is the detuning of the  fluctuators, and 
\begin{equation}
g_{n}=E_{J}j_{xn}\sqrt{\hbar/(2C_{0}\omega_{c})}\sin(2e\Phi_s/\hbar)
\end{equation} 
is the coupling constant. Note that coupling constant for other coupling mechanisms such as dielectric coupling between the  fluctuators and the resonator can be derived similarly~\cite{TianPRL2007}.
The decay of the resonator is modeled by its coupling to a bath of modes described by  the annihilation operator $a_{k}$ with frequency $\omega_{k}$ and coupling constants $c_{k}$. The decay rate is given by $\kappa=\pi\sum c_{k}^{2}\delta(\omega-\omega_{k})$~\cite{WallsBook1994}. The Hamiltonian $H_{t}$ describes a typical cavity QED system between the  fluctuators and the junction resonator~\cite{HoodScience}. 

Note that the driving on the resonator generates a time-dependent oscillation in the phase variable with the amplitude $\delta\Phi_{d}=\delta I_{c}/C_0\omega_c^2$. To keep the nonlinear term in the Josephson energy to be small, the oscillation amplitude needs to be small, e.g. $|2e \delta\Phi_{d}/\hbar|<0.1$. With $1/L\sim 4e^2E_{J}/\hbar^2$ and typical parameters $E_{J}\sim2\pi\times100\,\textrm{GHz}$ and $C_{0}\sim10^{-12}\,\textrm{pF}$, we estimate that the driving amplitude is bounded by $\epsilon\le2\pi\times 1 \textrm{GHz}$. 

\section{the dispersive regime\label{dispersive}} 
In this work, we study the quantum logic operations in the dispersive regime where the coupling $g_{n}$ is much weaker than the detuning between the  fluctuators and the resonator: $g_{n}\ll |\Delta_{nc}|$ with $ \Delta_{nc}\equiv\Delta_{n}-\Delta_{c}$. In this regime, we can apply 
the following unitary transformation~\cite{BlaisPRA2004} 
\begin{equation}
U=e^{-\epsilon(a-a^{\dagger})/\Delta_{c}}\prod_{n}e^{-g_{n}(a^{\dagger}\sigma_{n-}-\sigma_{n+}a)/\Delta_{nc}},\label{U}
\end{equation} 
to the system. After the transformation, the Hamiltonian becomes $\widetilde{H}_{t}=UH_{t}U^{\dagger}$ with $\widetilde{H}_{t}=H_{c}+\widetilde{H}_{1}+\widetilde{H}_{x}$ to the second order of $g_{n}/\Delta_{nc}$. 
The Hamiltonian is now divided into three parts: a Hamiltonian for the resonator $H_c$, an effective Hamiltonian for the  fluctuators $\widetilde{H}_{1}$, and  a small residual coupling between the  fluctuators and the resonator $\widetilde{H}_{x}$.  

The Hamiltonian $\widetilde{H}_{1}$ can be written as 
\begin{eqnarray}
     \widetilde{H}_{1} &=& \sum_{n}\left[ \frac{\hbar\widetilde{\Delta}_{n}}{2}\sigma_{nz}+\frac{\Omega_{nx}}{2}\sigma_{nx} \right] +H_{int}+\widetilde{H}_{k} 
     \label{eq:H1eff}\\ 
     H_{int} &=& \sum\lambda_{mn}(\sigma_{n+}\sigma_{m-}+\sigma_{m+}\sigma_{n-})/2\label{Hint} \\
     \widetilde{H}_{\kappa} &=& \sum_{n,k}( g_{n}c_{k}/\Delta_{nc})(\sigma_{n+}a_{k}+a_{k}^{\dagger}\sigma_{n-})
\end{eqnarray}
which includes the effective single qubit terms, an exchange-like interaction $H_{int}$, and 
an induced coupling to the bath modes of the resonator $\widetilde{H}_{\kappa}$. We derive the  
detuning for the single qubits as
\begin{equation} 
\widetilde{\Delta}_{n}=\Delta_{n}+(g_{n}^{2}/\Delta_{nc}) (1 - 2\epsilon/\Delta_{c})\label{Delta}
\end{equation} 
and the Rabi frequency as 
\begin{equation}
\Omega_{nx}=2\epsilon g_{n}/\Delta_{nc}.
\end{equation}
The coupling constant in the exchange-like interaction can  be derived as
\begin{equation}
   \lambda_{mn}=g_{m}g_{n}(\Delta_{mc}+\Delta_{nc})/(\Delta_{mc}\Delta_{nc}). \label{lambda}
\end{equation}
In the following section, we will study the implementation of the quantum logic gates with the  Hamiltonian $\widetilde{H}_{1}$. 

The residual coupling $\widetilde{H}_{x}$ can be written as 
\begin{equation}
      \widetilde{H}_{x}=\sum_{n}\frac{g_{n}^{2}}{\Delta_{nc}}\sigma_{nz}
         \left[ a^{\dagger}a+ \epsilon \left(\frac{\Delta_{c} - 2\Delta_{nc}}{2\Delta_{nc} \Delta_{c}}\right)  (a+a^{\dagger}) \right]  \label{eq:Hx}
\end{equation}
where the first term is the Stark shift for the resonator and
the second term is a coupling to the resonator amplitude originated from the finite driving amplitude.  Because of the amplitude shift in the
unitary transformation in Eq.~(\ref{U}), the average occupation of the resonator is now zero  with
$\langle a^\dagger a\rangle \approx 0$. Hence, the first term has a small effect on the  fluctuators during the quantum operations.  The second term can induce a small modification to the coupling constant $\lambda_{mn}$ in the effective interaction in Eq.~(\ref{Hint}) which will be studied in detail below.

\section{quantum logic gates\label{gates}} 
Universal quantum gates can be performed by controlling the effective Hamiltonian $\widetilde{H}_{1}$.  Here, we present the scheme for the single-qubit and two-qubit gates with typical parameters from superconducting Josephson junction resonators. 

\begin{figure}
\includegraphics[clip,width=8cm,clip]{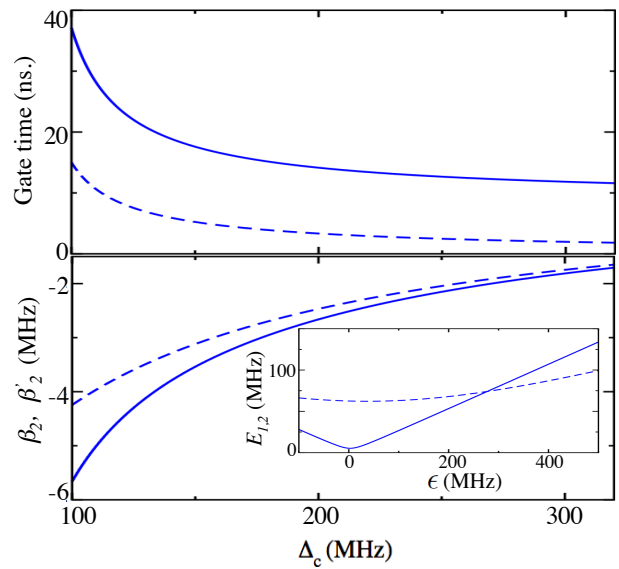}
\caption{\label{fig2} Upper plot: gate times for Hadamard gate (solid curve)
and spin flip gate (dashed curve) versus detuning $\Delta_{c}$.
Lower plot: effective couplings $\beta_{2}$ and
$\beta_{2}^{\prime}$ versus $\Delta_{c}$ (main plot); energy $E_{1}$
(solid curve) and $E_{2}$ (dashed curve) versus $\epsilon$ at $\Delta_{c}=300\,\textrm{MHz}$ (inset). }
\end{figure}
Single-qubit gates on a chosen TLS (e.g. $n=1$) can be performed by adjusting the frequency of the driving source to be close to the frequency of this TLS. By adjusting the driving amplitude and the detuning of the resonator, the effective qubit parameters $\widetilde{\Delta}_{1}$ and $\Omega_{1x}$ can be adjusted in a wide range. For a given detuning $\Delta_{c}$, one can adjust the driving amplitude $\epsilon$ in Eq. (\ref{Delta}) to have 
$\widetilde{\Delta}_{1}=0$ and obtain the spin flip gate $X$. The gate time can be found to be $\tau_{g} = \pi \Delta_{nc}/2\epsilon g_{n}$ at the chosen $\epsilon$.  One can also adjust $\epsilon$ to have $\widetilde{\Delta}_{1}=\Omega_{1x}$ to implement the Hadamard gate $H$. Here, the driving amplitude is chosen to be
\begin{equation}
\epsilon=\frac{(\Delta_c\Delta_{nc}+g_n^2)\Delta_c}{2(\Delta_c+g_n)g_n}
\end{equation}
with the gate time $\tau_{g} = \pi \Delta_{nc}/2\sqrt{2}\epsilon g_{n}$. 
In Fig.~\ref{fig2}, we plot the gate times of the Hadamard gate and spin flip gate with the parameters $\Delta_{1}=2\pi\times40\,\textrm{MHz}$ and
$g_{1}=2\pi\times40\,{\rm \textrm{MHz}}$. It can be shown that gate times on the order of $10\,\textrm{ns.}$ can be achieved. 
\begin{table}
\begin{tabular}{|c|c|c|c|c|}
\hline 
 & $\Delta_{c}$ ($2\pi\times$MHz) & $\epsilon$ ($2\pi\times$MHz) & $\Omega_{1x}$ ($2\pi\times$MHz) & time (ns.) \\
 \hline
X & 120 & -60 & 60 & 8.3 \\
\hline
H & 160 & -32 & 21.3 & 16.6 \\
\hline
\end{tabular}
\caption{\label{tab1}Example parameters for implementing the spin flip gate (X) and the Hadamard gate (H).}
\end{table}
In Table \ref{tab1}, we list two sets of gate parameters at the detunings $\Delta_{c}=2\pi\times160\,\textrm{MHz}$ for the Hadamard gate and $\Delta_{c}=2\pi\times120\,\textrm{MHz}$ for the spin flip gate respectively as an example. The corresponding driving amplitudes are listed in the table. For  fluctuators not involved in the gate (e.g. $n=2$), we have $\hbar\widetilde{\Delta}_{n}\gg\Omega_{nx}$ due to the fact that the  fluctuators are well separated in energy.  Hence, the gate operation only induces a dynamic phase to these  fluctuators. Meanwhile, the effective interaction between the  fluctuators is also prevented from generating controlled phases due to the off-resonance condition $\lambda_{1n}\ll|\widetilde{\Delta}_{1}-\widetilde{\Delta}_{n}|$.

Two-qubit gates can be performed via the effective exchange-like coupling in Eq.~(\ref{eq:H1eff}). 
This coupling can generate SWAP gate and $\sqrt{\textrm{SWAP}}$-like controlled gates when the two qubits are near resonance with $\widetilde{\Delta}_{m}-\widetilde{\Delta}_{n}\sim0$ \cite{BlaisPRA2007}. However, as noted above, the TLS  fluctuators in our system are usually far off-resonance from each other due to the large energy separations and are also hard to be manipulated individually. We will now show that two  fluctuators can be brought into effective resonance by controlling the resonator mode, and hence high fidelity two-qubit gates can be performed.  We first rewrite the single TLS energy in $\widetilde{H}_{1}$ as $\sum_{n}(E_{n}/2)\bar{\sigma}_{nz}$ where $E_{n}^2= \widetilde{\Delta}_{n}^{2}+\Omega_{nx}^{2}$ and 
\begin{equation}
\bar{\sigma}_{nz} = \cos\theta_{n}\sigma_{nz}+\sin\theta_{n}\sigma_{nx}
\end{equation} 
with $\cos\theta_{n}=\widetilde{\Delta}_{n}/E_{n}$ and $\sin\theta_{n}=\Omega_{nx}/E_{n}$.  Note $\bar{\sigma}_{nx}$ and $\bar{\sigma}_{ny}$ can be defined similarly. As the driving amplitude increases, the Rabi frequency
increases accordingly and the detuning $\widetilde{\Delta}_{n}$ will be affected.
As plotted in the inset of Fig.~\ref{fig2}, a driving amplitude can be found 
where $E_n$ is the same for the two  fluctuators. Denoting these  fluctuators as $n=1$ and $n=2$, we have $E_{1}=E_{2}$ at this point,  and they are now in resonance. In the interaction frame, the effective Hamiltonian for these two  fluctuators then becomes
\begin{equation}
H_{12}=\beta_{1}\bar{\sigma}_{1z}\bar{\sigma}_{2z}+\beta_{2}(\bar{\sigma}_{1+}\bar{\sigma}_{2-}+\bar{\sigma}_{1-}\bar{\sigma}_{2+})\label{Hrot}
\end{equation}
with the coefficients
\begin{eqnarray}
\beta_{1} &=& (\lambda_{12}\Omega_{1}\Omega_{2}/4E_{1}^{2}) \\
\beta_{2} &=& (\lambda_{12}/4)(1+ \widetilde{\Delta}_{1}\widetilde{\Delta}_{2}/E_{1}^{2})
\end{eqnarray}
respectively. 
Meanwhile, at large driving amplitude with $\epsilon\sim\Delta_{c}$, the second term in the residual coupling $\widetilde{H}_{x}$ in Eq. (\ref{eq:Hx}) induces virtual transitions between the resonator and the  fluctuators, which modifies the coupling constant $\beta_{2} $ to become $\beta_{2}^\prime$. Denoting the second term in Eq.(\ref{eq:Hx}) as $f_{n}\sigma_{nz}(a+a^\dagger)$, we can derive
\begin{equation}   \beta_{2}^{\prime}=\beta_{2}+\frac{f_{1}f_{2}(E_{1}+E_{2}-2\Delta_{c})}{2(E_{1}-\Delta_{c})(E_{2}-\Delta_{c})}. 
\end{equation}
In Fig.~\ref{fig2}, we plot both the couplings $\beta_{2}$ and $\beta_{2}^{\prime}$ versus the resonator detuning $\Delta_c$ for comparison. A small but finite modification of the coupling coefficient can be seen. Note when $|E_{1,2}-\Delta_{c}|\sim\lambda_{12}$, the second term in $\widetilde{H}_{x}$ induces real transitions between the  fluctuators and the resonator, which can seriously affect the gate operations and should be avoided when designing the gates. 

Two-qubit gates of the form of $S_{0}=\exp(-iH_{12}^{rot}t)$ can now be performed in the rotating frame. A SWAP gate has the gate time $\tau_{g}=\pi/2|\beta_{2}^{\prime}|$. The $\beta_{1}$
term in Eq. (\ref{Hrot}) contributes only phase factors in the computational basis to the gate operation. With the following parameters: $\Delta_{1}=0$, $\Delta_{2}=-2\pi\times60\,{\rm {MHz}}$, $g_{1}=2\pi\times40\,{\rm {MHz}}$, $g_{2}=2\pi\times30\,{\rm {MHz}}$,
and $\Delta_{c}=2\pi\times300\,{\rm {MHz}}$, we find that $E_{1}=E_{2}=2\pi\times74.0\,\textrm{MHz}$ at $\epsilon=2\pi\times277.2\,{\rm {MHz}}$. 
Here, $\beta_{2}^{\prime}=-2\pi\times1.8\,{\rm {MHz}}$ and the SWAP gate can be performed with a gate time of $\tau_{g}=137.9\,\textrm{ns.}$. 

We note that controlled quantum logic gates can also be performed using the Cirac-Zoller gate which was first studied in ion trap quantum
computing~\cite{LeibfriedRMP}.  This gate includes three steps: a swap gate between the first TLS and the resonator, a conditional phase gate between the
resonator and the second TLS, and another swap gate between the first
TLS and the resonator. The swap gate in the first and third steps can be implemented by choosing  $\Delta_{c}=\Delta_{1}$ and $\epsilon=0$, and the gate time is  $\pi/2g_{1}$ which is of the order of a few nanoseconds. In the conditional phase gate, the driving frequency is close
to the chosen qubit but is still in the dispersive regime. The Stark shift $(g_{2}^{2}/\Delta_{2c})\sigma_{2z}a^{\dagger}a$ generates a conditioned phase shift on this TLS when the resonator is in state $|1\rangle$. The gate time is $t_{cg}=\pi\Delta_{2c}/2g_{2}^{2}$. At $g_{2}=2\pi\times30\,\textrm{MHz}$ and $\Delta_{2c}=2\pi\times120\,\textrm{MHz}$, $t_{cg}\sim30\,\textrm{ns.}$. Note that the first TLS is subject to stronger decoherence during the swap operation due to its near-resonance coupling with the resonator, as will be discussed below.

\section{Decoherence\label{decoherence}} 
The intrinsic decoherence of the TLS  fluctuators is very slow and can be ignored during the gate operation. However, the coupling between the resonator and the 
 fluctuators induces decoherence that cannot be neglected.  In the dispersive regime, the decoherence rate can be calculated from the noise term $\widetilde{H}_{\kappa}$ in Eq.(\ref{eq:H1eff}). It can be shown that the decoherence rate is on the order of $\tau_{d}^{-1}\sim g_{n}^{2}\kappa/\Delta_{nc}^{2}$ during the quantum logic gates with $\tau_d^{-1}\ll\kappa$ in the dispersive regime.  In contrast, the decoherence rate during the swap operation in the Cirac-Zoller gate is $\tau_{d}^{-1}\sim\kappa/2$ which is much faster than the decoherence rate in the dispersive regime.   In Table \ref{tab2}, we list the gate times, the decoherence rates, and the ratios of gate times to decoherence
times for the gates discussed above at $\kappa=4\,\textrm{MHz}$. 
\begin{table}
\begin{tabular}{|c|c|c|c|}
\hline 
 & 1-Qubit & 2-Qubit & swap op. \\
\hline
$\tau_{g}^{}$  & $\pi\Delta_{nc}/g_{n}\epsilon$ & $\pi/2|\beta_{2}^{\prime}|$ & $\pi/2g_{n}$ \\
\hline 
$\tau_{d}^{-1}$ & $g_{n}^2\kappa/\Delta_{nc}^{2}$ & $g_{n}^2\kappa /\Delta_{nc}^{2}$ & $\kappa/2$ \\
\hline 
$\tau_{g}/\tau_{d}$ & $10^{-3}$ & 0.01 & $0.02$ \\ \hline
\end{tabular}
\caption{\label{tab2}Gate times $\tau_{g}$, decoherence rate $\tau_{d}^{-1}$, and the ratios $\tau_{g}/\tau_{d}$ for single-qubit and two-qubit gates.  The column labelled as ``2-Qubit'' is for the two-qubit gates studied in our scheme and the column labelled as ``swap op.'' is for the swap operation in the Cirac-Zoller gate~\cite{LeibfriedRMP}.}
\end{table}

One can estimate the gate fidelity approximately as 
\begin{equation}
F=e^{-\tau_{g}/\tau_{d}}\approx1-\tau_{g}/\tau_{d}
\end{equation} 
which depends on the ratio between the gate time and the decoherence time. 
Using the parameters given above, we can estimate the ratio $\tau_g/\tau_d$. In Table \ref{tab2}, it is shown that $\tau_{g}/\tau_{d}\approx 10^{-2}$ for the two-qubit gate in our scheme and  $\tau_{g}/\tau_{d}\approx 2\times10^{-2}$ for the SWAP operation in the Cirac-Zoller gate at 
the damping rate $\kappa=4\,\textrm{MHz}$.  The gate fidelities can thus reach $0.99$ with our protocols. This indicates that the Josephson junction resonator is an effective tool in generating high fidelity quantum operations even at a quality factor ($Q=\omega_c/\kappa$) of only $Q=7800$ and hence can be used to demonstrate quantum coherence in the TLS  fluctuators.  

The term $\sigma_{nz}(a+a^{\dagger})$ in the residual coupling
$\widetilde{H}_{x}$ in Eq.~(\ref{eq:Hx}) can induce additional decoherence.
Here, quantum fluctuations of the resonator mode cause extra
noise even at zero resonator amplitude. The spectrum of the quantum fluctuations can be derived as 
\begin{equation}
\langle aa^{\dagger}\rangle_{\omega}=((\omega-\omega_{c})^{2}+\kappa^{2}/4)^{-1}\kappa,
\end{equation}
from which the decoherence rate can be derived as  
\begin{equation}
\tau_{d}^{-1}\sim(\epsilon^{2}g_{n}^{4}/\Delta_{nc}^{6}){\rm \kappa}.
\end{equation} 
As can be seen, the decoherence rate is to the $4$th order of $g_{n}/\Delta_{nc}$ when
the driving amplitude $\epsilon$ is comparable with the detunings. With the above
parameters, we find that $\tau_{d}^{-1}\approx 1\,\mbox{kHz}$ which can be neglected during the gate operations.

\begin{figure}
\includegraphics[width=8cm,clip]{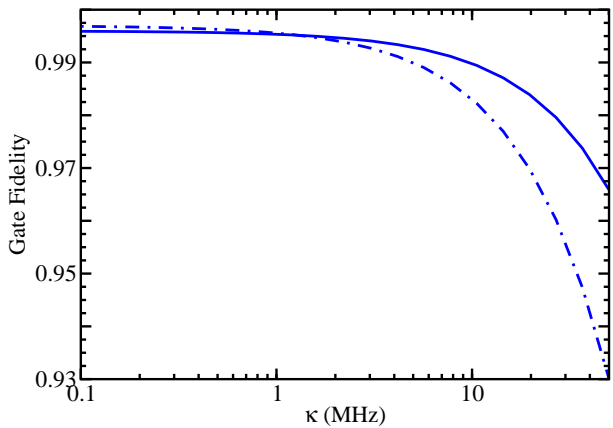}
\caption{\label{fig3}Gate fidelity versus damping rate $\kappa$ by numerical simulation. Solid curve: SWAP gate and dash-dot curve: Hadamard gate.  Two  fluctuators are included in the simulation for both gates.  For the Hadamard gate, we use the parameters $\Delta_{2}=-2\pi\times80\,{\rm \textrm{MHz}}$ and $g_{2}=2\pi\times30\,{\rm \textrm{MHz}}$ for TLS $n=2$.  }
\end{figure}
To study the effectiveness of the quantum logic gates, we perform numerical
simulations on the gate operations using the full Hamiltonian $\widetilde{H}_{t}$.  
The decay of the resonator is simulated using the Lindblad master equation~\cite{WallsBook1994}.  This simulation includes both the effect of the residual coupling in Eq. (\ref{eq:Hx}) and the effect of the resonator decoherence.  We calculate the fidelities of the Hadamard and SWAP gates using the method prescribed by Nielsen's formula for the gate fidelity~\cite{NielsenPLA2002} over a wide range of resonator decay rate. The results are plotted in Fig.~\ref{fig3}. This simulation shows that the fidelity can be higher than $F=0.99$ for  
$\kappa\le 5\,\mbox{MHz}$ for single-qubit and two-qubits gates, which also agrees with our estimations above.   Hence, at moderate decay rate for the Josephson junction resonator, high fidelity quantum logic gates can be achieved.  

\section{Discussions \label{discussions}} 
%readout
As demonstrated in recent works, the superconducting resonator can be used to detect the states of qubits or fluctuators~\cite{BlaisPRA2004,TianPRL2007}. In the strong damping regime, the amplitude of the superconducting resonator adiabatically follows the dynamics of the  TLS fluctuators, as was shown in our previous work~\cite{TianPRL2007}. A phase sensitive measurement of the resonator can hence provide a direct measurement of the TLS states. In the moderate damping regime where the damping rate is weaker than the coupling constant, a measurement of a TLS can be performed by adjusting the resonator frequency to be in the vicinity of this TLS, but with the condition $g_{n}\ll |\Delta_{nc}|$.  Here, the Stark shift resulting from this TLS is much stronger than that from other qubits. A measurement of the resonances in the transmission spectrum of the resonator can then provide a measurement of the qubit state~\cite{BlaisPRA2004}. Such measurements can be realized with current electronics where the resonator and the driving can be adjusted and switched on-and-off in nanoseconds, much faster than the decoherence time of the  fluctuators. 

%scalability
Our scheme can also be extended to TLS  fluctuators inside different junctions. Because the wavelength of the microwave mode of the superconducting resonator is much longer than the dimension of this circuit,  fluctuators in different junctions can couple to the same resonator mode when the junctions are connected by a superconducting loop. As is illustrated in Fig.~\ref{fig1} (b), two junctions are connected to the central superconducting island labelled by $\Phi$ that is associated with the junction resonator.  It can be shown that effective coupling between  fluctuators in the two junctions can be derived exactly as described by Eq.~(\ref{eq:H1eff}). 
In this configuration, quantum logic gates can be performed with essentially the same approach as was presented above.  This circuit can also be extended to include multiple junctions. This system is thus intrinsically ``scalable'' where  fluctuators in multiple junctions couple nonlocally. Note that  
the frequency of the superconducting resonator is determined by the total capacitance, the total effective Josephson energy, and the inductance in the circuit, and will decrease as the number of junctions increases. This can set a limit on the number of junctions that can be connected into the circuit. 

%discussion
TLS  fluctuators have been studied for a long time. Previously, the  fluctuators are often considered as a  source of decoherence in superconducting qubits, causing the so-called $1/f$ noise.  In this work, we focused on studying  the coherent manipulation of the  fluctuators which can provide insights about the dynamics, the coupling mechanism, and the relaxation of the  fluctuators in superconducting devices.  Although the success in implementing universal quantum logic gates makes the TLS  fluctuators potential candidate for quantum computing, we want to emphasize that the main aim of this work is to provide a practical scheme to demonstrate the coherence behavior of the  fluctuators. Our scheme can be useful for current experiments that investigate the coupling between the  fluctuators and superconducting resonators~\cite{SimmondsPRL,KimRecentExp,ZagoskinPRL,Osborn2007,TianPRL2007,Cicak2008}. 

\section{Conclusions \label{conclusions}} 
To conclude, we have shown that universal quantum logic gates can be implemented on spurious  TLS fluctuators via the coupling between the  fluctuators and the Josephson junction resonator.  Taking into account the full Hamiltonian of the coupled system and the effect of the noise, our numerical simulation of the quantum operations showed that quantum logic gates can be performed with high fidelity even at resonator decay rates of a few megahertz. We have used practical parameters for the junction resonators and the  fluctuators in this study. Our work hence indicates that quantum coherence and quantum manipulation of TLS  fluctuators can be readily demonstrated. The results here can be generalized to other types of superconducting resonators which are explored in recent experiments~\cite{Osborn2007}.

\end{document}